\documentclass[%
 reprint,
superscriptaddress,
nofootinbib,
 amsmath,amssymb,
 aps,
pra,
]{revtex4-1}

\usepackage{graphicx}
\usepackage{dcolumn}
\usepackage{bm}

\usepackage[utf8]{inputenc}
\usepackage{xcolor}
\usepackage{graphicx}
\usepackage{amsmath}
\usepackage[section]{placeins}
\usepackage{dblfloatfix}
\usepackage{appendix}
\usepackage{amsmath}
\usepackage{breqn}
\usepackage{tabu}
\usepackage[normalem]{ulem}
\usepackage{hyperref}

\def\ket#1{\left | {#1} \right \rangle}
\def\bra#1{\left \langle {#1} \right |}
\def\braket#1#2{\left \langle {#1} | {#2} \right \rangle}
\def\expec#1{\left \langle #1 \right \rangle}

\begin{document}


\title{Experimental Comparison of Bohm-like Theories with Different Primary Ontologies}

\author{Arthur O. T. Pang}
\email{arthur.pang@mail.utoronto.ca}
\author{Hugo Ferretti}
\author{Noah Lupu-Gladstein}
\author{Weng-Kian Tham}
\author{Aharon Brodutch}
\email{brodutch@physics.utoronto.ca}
\affiliation{Department of Physics  and  Centre for Quantum Information   Quantum Control University of Toronto$,$ 60 St George St$,$ Toronto$,$ Ontario$,$ M5S 1A7$,$ Canada}
\author{Kent Bonsma-Fisher}
\email{kent.bonsma-fisher@nrc-cnrc.gc.ca}
\affiliation{Department of Physics  and  Centre for Quantum Information   Quantum Control University of Toronto$,$ 60 St George St$,$ Toronto$,$ Ontario$,$ M5S 1A7$,$ Canada}
\affiliation{National Research Council of Canada$,$ 100 Sussex Dr$,$ Ottawa$,$ Ontario$,$ K1A 0R6$,$ Canada}
\author{J. E. Sipe}
\email{sipe@physics.utoronto.ca}
\affiliation{Department of Physics  and  Centre for Quantum Information   Quantum Control University of Toronto$,$ 60 St George St$,$ Toronto$,$ Ontario$,$ M5S 1A7$,$ Canada}
\author{Aephraim M. Steinberg}
\email{steinberg@physics.utoronto.ca}
\affiliation{Department of Physics  and  Centre for Quantum Information   Quantum Control University of Toronto$,$ 60 St George St$,$ Toronto$,$ Ontario$,$ M5S 1A7$,$ Canada}
\affiliation{Canadian Institute for Advanced Research$,$ Toronto$,$ Ontario$,$ M5G 1M1$,$ Canada}

\date{\today}

\begin{abstract}
The de Broglie-Bohm theory is a hidden-variable interpretation of quantum mechanics which involves particles moving through space along deterministic trajectories. This theory singles out position as the primary ontological variable. Mathematically, it is possible to construct a similar theory where particles are moving through momentum-space, and momentum is singled out as the primary ontological variable. In this paper,  we construct the putative particle trajectories for a two-slit experiment in both the position and momentum-space theories by simulating particle dynamics with coherent light. Using a method for constructing trajectories in the primary and non-primary spaces, we compare the phase-space dynamics offered by the two theories and show that they do not agree. This contradictory behaviour underscores the difficulty of selecting one picture of reality from the infinite number of possibilities offered by Bohm-like theories.
\end{abstract}

\maketitle

\section{Introduction}
Bohm's hidden-variable interpretation of quantum mechanics \cite{Bohm1952,PhysRev.85.180}, also known as Bohmian mechanics or de Broglie-Bohm theory \cite{deBroglie1927, Holland_book}, is an alternative formulation of quantum mechanics with a clear deterministic ontology, and experimental predictions that match those of quantum theory. 
The theory continues to attract attention as a provocative demonstration that a deterministic, no-collapse interpretation of quantum mechanics is possible \cite{Gambetta2004,Philippidis1979,Kocsis1170,Mahler2016,Zhou2017,Xiao2017,Xiao2019}. 
As is the case in classical physics, Bohmian particles have well defined properties at all times. 
In Bohmian theory all properties can be determined from the particle's actual position and the guiding wave, giving position special ontological significance.  
Wiseman~\cite{Wiseman2007} showed that it is possible to experimentally extract the velocities attributed to Bohmian particles by taking conditional averages of weak measurements on an ensemble of post-selected systems.
Extending his ideas, some of the present authors and others were recently able to construct the putative  Bohmian trajectories in various two-slit experiments~\cite{Kocsis1170,Mahler2016,Xiao2017,Xiao2019}.

The choice of position as the primary ontological variable introduces an asymmetry which is foreign to both classical and quantum mechanics. 
In classical Hamiltonian mechanics, position and momentum act as the canonical phase space variables and are both equally important in formulating the theory, and in orthodox quantum mechanics position and momentum are treated on equal footing. So the importance placed on position in the Bohmian approach was one of the main criticisms of Bohm's work by the pioneers of quantum theory
~\cite{Proof_not_proving}.  
Shortly after Bohm's paper appeared, Epstein~\cite{Epstein1953a,Epstein1953b} pointed out that there is nothing inherent in the formulation that requires position to be the primary ontological variable, and that other possible choices can lead to other results - i.e. different ontological descriptions - while still yielding experimental predictions identical to those of quantum theory.
 
In this paper, we demonstrate how different choices of the primary ontological variable can lead to qualitatively different trajectories. Using light in a double-slit setup to simulate the mechanics of massive particles \cite{Flack2018}, and building on a previously demonstrated approach \cite{Mahler2016,Kocsis1170},  we construct the trajectories in both Bohm's theory (which we refer to as $x$-Bohm) and an alternative theory in which momentum is the primary ontological variable ($p$-Bohm). 
The differences between the trajectories in the two theories illustrate why the results of previous experiments  \cite{Mahler2016,Kocsis1170} should be understood as specific instances of the many possible ontological descriptions of the same system. 
Selecting one out of the multitude of possible theories -- with their conflicting phase-space dynamics of the state of the system -- requires supplementary assumptions or assertions \cite{Wiseman2007,Gambetta2004,QuantumTheoryCrossroads}, emphasizing one of the features in Bohm's approach that some would consider a weakness.

We begin in Sec.~\ref{sec:BohmTheory} by describing some of the basic features of the $x$-Bohm and $p$-Bohm theories, and the method for constructing trajectories through a sequence of weak and strong measurements. 
In Sec.~\ref{sec:experiment} we lay out the details of our experiment, including the specifics of the lens system and the weak measurement procedure.  
The results of the experiments, including plots of the trajectories and phase space snapshots at the near and far field are presented in Sec.~\ref{sec:results} for both the  $x$-Bohm and $p$-Bohm theories. 
The implications of our results are discussed in Sec.~\ref{sec:discussion}.

\section{Bohmian theory}

\label{sec:BohmTheory} Contrary to classical mechanics, which allows for the deterministic prediction of the motion of particles, quantum mechanics only offers statistical predictions for the results of measurements.
Yet in 1952 David Bohm introduced~\cite{Bohm1952,PhysRev.85.180} a deterministic dynamical theory that its advocates argue provides an underlying description more fundamental than quantum mechanics \cite{undivided,QuantumTheoryCrossroads}.
In his generalization and extension of earlier ideas by de Broglie \cite{QuantumTheoryCrossroads}, the positions of particles play the role of hidden-variables; their motion is characterized by well-defined trajectories, as the particles are ``guided'' by the Schr\"odinger wave. 
In this approach, position variables, together with the Schr\"odinger wave, have a special significance as the primary ontological variables; the momenta of particles simply follow from their velocities, which are determined by the gradient of the Schr\"odinger wave at the positions of the particles.
The symmetry of position and momentum that characterizes orthodox quantum mechanics is broken, with position variables more fundamental than momentum variables.

Shortly after Bohm's work appeared, Epstein~\cite{Epstein1953a,Epstein1953b} noted that different choices of the primary ontological variable can lead to different theories. 
In particular, one could work with the momentum representation of the wave function and build a theory where particles are characterized fundamentally by their momenta\footnote{The possibility of a velocity-based theory had already been raised by Pauli at the 1927 Solvay conference in response to de Broglie's pilot wave theory~\cite{QuantumTheoryCrossroads,Bohm1952}.}. 
In contrast to Bohm's original theory, which we refer to as ``$x$-Bohm,'' in Epstein's proposal, which we refer to as a ``$p$-Bohm'' theory, it is momentum that has primary ontological status. In his reply to Epstein~\cite{Bohm1953}, Bohm pointed out technical difficulties in implementing a ``$p$-Bohm'' approach when the Hamiltonian involved the Coulomb potential. 
But he also argued that an ``$x$-Bohm'' approach, where particle position and the coordinate representation of the wave function are the primary ontological variables, seemed more favored because ``in all fields other than the quantum theory, space and time have thus far stood out as the natural frame for the description of the progress of physical phenomena.''~\cite{Bohm1953}
Nevertheless, alternate approaches were developed further a few decades later by Bohm and his collaborators~\cite{Brown2000}, and a general framework for such theories was discussed by Holland~\cite{Holland_book,Holland1993,Holland1998} and others~\cite{Struyve_2008,Struyve_2010}. 

In the rest of this section we sketch both $x$-Bohm and $p$-Bohm theory, discuss the trajectories that follow from each, and show how -- under the assumption that one of the theories is correct -- its associated trajectories can be revealed by weak measurements. 
We begin with trajectories of the primary ontological variable of the particles -- position for $x$-Bohm and momentum for $p$-Bohm -- and then turn to the trajectories that can be associated with non-primary variables.
This allows us to compare the two theories by contrasting their predictions for trajectories in the same space. 
We focus on the one-dimensional motion of a single particle, where the classical Hamiltonian as a function of position and momentum is $H(x,p)$, and denote the coordinate wave function by $\psi(x,t)$ and the momentum wave function by $\tilde{\psi}(p,t)$.
The Schr\"odinger equations for these two wave functions are 
\begin{align}
\nonumber \\
i\hbar\frac{\partial}{\partial t}\psi(x,t)= & H\left(x,-i\hbar\frac{\partial}{\partial x}\right)\psi(x,t)\label{eq:SEqnx}\\
i\hbar\frac{\partial}{\partial t}\tilde{\psi}(p,t)= & H\left(i\hbar\frac{\partial}{\partial p},p\right)\tilde{\psi}(p,t).\label{eq:SEqnp}
\end{align}

\subsection{Position Ontological Bohmian Theory ($x$-Bohm)}
\label{sec:xbohm_intro} 

In Bohm's original theory~\cite{Bohm1952,PhysRev.85.180}, the particle's position and the wave function $\psi(x,t)$ constitute the objectively real elements from which all other properties can be derived\footnote{For an $N$-particle system the wave function is a function over the $3N$-dimensional configuration space of the system, and since the wave function is granted ontological significance that configuration space must be taken as the underlying arena of reality; the wave function and a point in this configuration space, identifying the positions of all $N$ particles, are best taken to identify the ontology of the theory.}. 
In describing an ensemble of experimental runs, at some initial time $(t=0)$ each particle is assumed to have a definite position according to a probability distribution function $|\psi(x,0)|^{2}$, and each particle is guided through space by the wave function. 
Writing $\psi\left(x,t\right)=R_{x}\left(x,t\right)\exp\left[iS_{x}\left(x,t\right)/\hbar\right]$, where $R_{x}\left(x,t\right)$ and $S_{x}\left(x,t\right)$ are real functions of position and time, for a Hamiltonian of the form $H(x,p)=p^{2}/2m+V(x)$ the guidance equation is 
\begin{equation}
v_{x}\left(x,t\right)=\frac{1}{2m}\frac{\partial S_{x}(x,t)}{\partial x},\label{eq:Bohm_velocity_a}
\end{equation}
and the trajectory for each particle is given by
\begin{equation}
\frac{dx\left(t\right)}{dt}=v_{x}\left(x\left(t\right),t\right).
\end{equation}
Since the expression (\ref{eq:Bohm_velocity_a}) for the velocity
$v_{x}(x,t)$ can also be written as~\cite{Holland_book}
\begin{align}
 & v_{x}\left(x,t\right)=\frac{j_{x}\left(x,t\right)}{\left|\psi\left(x,t\right)\right|^{2}},
 \label{eq:Bohm_velocity_again}
\end{align}
where $j_{x}(x,t)$ is the usual
probability current density of orthodox quantum mechanics, 
\begin{align}
 & j_{x}\left(x,t\right)=\frac{\hbar}{2mi}\left(\psi^{*}\left(x,t\right)\frac{\partial\psi\left(x,t\right)}{\partial x}-\psi\left(x,t\right)\frac{\partial\psi^{*}\left(x,t\right)}{\partial x}\right),\label{eq:current_density}
\end{align}
it follows that as the particles in the ensemble move, and as $\psi(x,t)$
evolves according to Schr\"odinger's equation (\ref{eq:SEqnx}), the
evolution of the distribution function characterizing the positions
of the particles follows the evolution of $\left|\psi(x,t)\right|^{2}$.

Although the Bohmian trajectories had been studied theoretically and discussed in the literature since 1952 (see, e.g., Philippidis et al.~\cite{Philippidis1979}), it seems it was not until Wiseman's work in 2007 \cite{Wiseman2007} that a strategy for identifying them experimentally was investigated. 
Wiseman pointed out that the expression (\ref{eq:Bohm_velocity_a}) for the velocity of a particle at $x$, which can be written as~\cite{Holland_book} 
\begin{equation}
v_{x}\left(x,t\right)=\frac{1}{m}\text{Re}\left[\frac{\left<x\right| \hat p \left|\psi\left(t\right)\right>}{\left<x|\psi\left(t\right)\right>}\right],
\label{eq:Bohm_velocity_operational}
\end{equation}
where $\hat p$ is the momentum operator ($\left\langle x| \hat p |x'\right\rangle=-i\hbar\partial\delta(x-x')/\partial x$), can be connected with the theory of weak measurements introduced by Aharonov, Albert, and Vaidman (AAV)~\cite{Aharonov1988}. 
Weak measurements are those with small back action and consequently high uncertainty, and Wiseman noted that the expression (\ref{eq:Bohm_velocity_operational}) corresponds to the operational prescription of a weak momentum measurement followed immediately by a strong (projective) position measurement.
The apparent simultaneous measurement of two conjugate variables respects the uncertainty relations since the momentum measurement is weak,
and consequently the measurement scenario must be repeated many times with the averaging done separately for every final value of position.
This fits neatly into the Bohmian perspective in general: Since all variables in the theory are uniquely determined by the primary ontological variable, it could be argued that ensemble averaging is justified as long as post-selection onto the primary ontological value for each experimental run is sufficiently accurate and the measurement back action from the weak measurement is sufficiently small.

Of course, the identification of the right-hand-side of (\ref{eq:Bohm_velocity_operational}) with a weak momentum measurement followed by a strong position measurement can be made operationally, independent of any proposed explanation of quantum mechanics in terms of a deeper theory. 
Nonetheless, the trajectories that are predicted by $x$-Bohm theory can be formally constructed from the results of weak measurements; this has been done by Kocsis et al.~\cite{Kocsis1170} for a single particle in a double-slit interferometer, and by Mahler et al.~\cite{Mahler2016} for entangled particles.
Advocates of $x$-Bohm theory then identify these constructed trajectories with trajectories that are held to really exist.

\subsection{Momentum Ontological Bohmian Theory ($p$-Bohm)}
\label{sec:pbohm_intro} 

In $p$-Bohm theory one adopts momentum as the primary ontological variable, and the fundamental dynamics take place in momentum-space. 
Here one relies on the momentum representation of the wave function $\tilde{\psi}(p,t)$, and for Hamiltonians of the form $H(x,p)=p^{2}/2m+V(x)$ there is no general expression for the time derivative $v_{p}(p,t)$ of the momentum of a Bohmian particle,
\begin{align}
 & \frac{dp\left(t\right)}{dt}=v_{p}\left(p\left(t\right),t\right),
\end{align}
which would be analogous to the corresponding expression (\ref{eq:Bohm_velocity_a}) for the time derivative $v_{x}(x,t)$ of the position of a Bohmian particle in $x$-Bohm theory. 
This can be traced to the fact that all such Hamiltonians exhibit the same dependence on $p$ but, depending on the potential, can have very different dependences on $x$; thus, while Schr\"odinger's equation in configuration space (\ref{eq:SEqnx}) involves only second derivatives with respect to $x$, Schr\"odinger's equation in momentum-space (\ref{eq:SEqnp}) can involve any number of derivatives with respect to $p$. 
Nonetheless, one can look for an expression for $v_{p}(p,t)$ analogous to the expression (\ref{eq:Bohm_velocity_again}) for $v_{x}(x,t)$, writing 
\begin{align}
 & v_{p}(p,t)=\frac{j_{p}(p,t)}{\left|\tilde{\psi}(p,t)\right|^{2}},\label{eq:momentum_velocity}
\end{align}
where $j_{p}(p,t)$ is a current density in momentum-space. 
In a one-dimensional problem it must satisfy 
\begin{align}
 & \frac{\partial j_{p}(p,t)}{\partial p}=-\frac{\partial}{\partial t}\left(\left|\tilde{\psi}(p,t)\right|^{2}\right),
 \label{eq:momentum_continuity}
\end{align}
and since the right-hand-side is determined by the Schr\"odinger equation in momentum-space (\ref{eq:SEqnp}), a unique $j_{p}(p,t)$ can be
identified, 
\begin{align}
 & j_{p}(p,t)=\frac{2}{\hbar}\int_{-\infty}^{p}\text{Im}\left(\tilde{\psi}^{*}(p',t)\left(V(i\hbar\frac{\partial}{\partial p'})\tilde{\psi}(p',t)\right)\right)dp',\label{eq:momentum current}
\end{align}
under the physically reasonable assumption that $j_{p}(p,t)\rightarrow0$ as $\left|p\right|\rightarrow\infty$~\cite{Struyve_2008}. 
The situation is more complicated in higher dimensions; in three dimensions, for example, the continuity equation for a momentum current density $\boldsymbol{j}_{p}(\boldsymbol{p},t)$ only restricts the divergence of $\boldsymbol{j}_{p}(\boldsymbol{p},t)$
and not its curl, and it is not immediately clear how it should be assigned. 
The range of possible choices for current densities in general de Broglie-Bohm theories, and the criteria one might want to apply in making a choice, have been investigated by Struyve and Valentini~\cite{Struyve_2008}. 
Our focus in this paper will be on free particles ($V(x)=0)$, where the distribution $\left|\tilde{\psi}(p,t)\right|^{2}$ of Bohmian particles in momentum-space is time independent, and from (\ref{eq:momentum_velocity},\ref{eq:momentum current}), and in agreement with physical intuition, we have $v_{p}(p,t)=0$.

\subsection{Trajectories in a non-primary space}
\label{sec:BohmTraj_nonstandard} 

Although a Bohmian theory is deterministic, with a well-defined evolution
 of dynamical variables once the initial conditions are specified, the phase
 space description of dynamics so useful in classical theories is at first
 sight not a natural one here.
 A Bohmian theory always identifies a primary ontological variable, which in 
 $x$-Bohm theory is particle position.
 Of course, in this theory one can multiply the velocity of the Bohmian
 particle – which is determined once the wave function and the particle
 position are specified – by the mass of the particle and so identify a
 particle momentum.
 But it has a diminished status in 
 $x$-Bohm theory compared to its role in classical dynamics, for the initial
 momentum of the particle cannot be specified independently of the initial
 position of the particle.
 The proper arena of dynamics seems to be configuration space, in which
 the particle position is a point and over which the wave function is defined,
 rather than phase space.

The situation is more drastic for 
 $p$-Bohm theory, especially in the case of a classically free particle that
 we consider here.
 Momentum is the primary ontological variable, the proper arena of dynamics is clearly the space associated with momentum, and for a free particle the momentum of the Bohmian particle
 does not change.
 The non-primary variable of particle position does not even seem to arise,
 and in this theory one can even wonder whether or not the question
 of what each particle is doing in configuration space is meaningful. Thus in both the $x$-Bohm and $p$-Bohm theories the significance of phase-space is unclear without proper treatment of the non-primary variables.

 In his consideration of the status of variables in Bohmian theories,
 Holland~\cite{Holland_book,Holland1993} suggested a strategy for identifying the values of variables other than the primary ontological variable, thus allowing in general for a visualization of the underlying dynamics in phase space.
For one-dimensional systems and in our notation, if we consider a ``$\xi$-Bohm theory,'' where here $\xi$ is an eigenvalue of a Hermitian operator $\hat \xi$ that we take to have continuous eigenvalues, the value $\omega$ of a continuous variable associated with a Hermitian operator $\hat \omega$ is taken to be 
\begin{align}
 & \omega_{\xi}(\xi,t)=\text{Re}\left[\frac{\left<\xi\right| \hat \omega \left|\psi\left(t\right)\right>}{\left<\xi|\psi\left(t\right)\right>}\right]\label{eq:Holland}
\end{align}
at time $t$, if the ket is $\left|\psi(t)\right\rangle $ and the primary ontological variable has value $\xi$. 
Holland did not take this suggestion to be at the level of a new postulate, and even considered
other approaches for some physical systems.
Nonetheless, the proposal has the physically comforting feature that the average of the values granted to a variable $\omega$ over an ensemble of Bohmian particles described by a $\xi$-Bohm theory does agree with the expectation value of the operator associated with that variable in the ket describing the ensemble, 
\begin{align}
 & \left\langle \psi(t)| \hat \omega |\psi(t)\right\rangle =\int \omega_{\xi}(\xi,t)\left|\left\langle \xi|\psi(t)\right\rangle \right|^{2}d\xi.
\end{align}

As an example, consider momentum in an $x$-Bohm theory. 
For a particle at position $x$ at time $t$, from (\ref{eq:Holland}) we see that the value of momentum that would be assigned is 
\begin{align}
 & p_{x}(x,t)=\text{Re}\left[\frac{\left<x\right| \hat p \left|\psi\left(t\right)\right>}{\left<x|\psi\left(t\right)\right>}\right].\label{eq:pinx}
\end{align}
Comparing with the $x$-Bohm expression (\ref{eq:Bohm_velocity_operational}) for $v_{x}(x,t)$, we find 
\begin{align}
 & p_{x}(x,t)=mv_{x}(x,t),\label{eq:pmv}
\end{align}
as would be physically expected. 
Yet we can now also assign evolving position variables to particles in a $p$-Bohm theory, for the prescription (\ref{eq:Holland}) gives 
\begin{align}
&x_p(p,t)=\text{Re}\left[\frac{\left<p\right| \hat x \left|\psi\left(t\right)\right>}{\left<p|\psi\left(t\right)\right>}\right]
 ,\label{eq:xinp}
\end{align}
and following $x_{p}(p,t)$ as $t$ advances allows us to assign a trajectory in real space to a Bohmian particle in $p$-Bohm theory
with momentum $p$.

Remarkably, Holland's prescription (\ref{eq:Holland}) is precisely that which operationally characterizes a weak $\hat \omega$ measurement followed by a strong $\hat \xi$ measurement.
Thus, just as velocities of particles in an $x$-Bohm theory can be constructed by weak momentum measurements followed by strong position measurements, so the positions of particles in a $p$-Bohm theory can be constructed by weak position measurements followed by strong
momentum measurements. 
And so we have a route to identifying trajectories
of Bohmian particles in ``non-primary'' spaces, by which we mean spaces associated with variables other than the primary ontological variable. 
This is done by first constructing the trajectories of the primary ontological variable, leading to an equation for $\xi(t)$ and then using Eq.~\eqref{eq:Holland} to construct the trajectory given by $\omega(\xi(t),t)$. 
We can experimentally construct trajectories in momentum-space for particles in an $x$-Bohm theory, as indeed has already been done implicitly in~\cite{Kocsis1170, Mahler2016} and explicitly in~\cite{Xiao2019}, both relying on (\ref{eq:pmv}). Additionally, we can also experimentally construct trajectories in position-space for particles in a $p$-Bohm theory. We will use these trajectories as basis to compare the phase-space dynamics of different Bohm-like theories, and will turn to this in the following sections.

\section{Experimental scheme}\label{sec:experiment}

In our experiment, we simulate the evolution of a massive particle under $x$-Bohm and $p$-Bohm theories with light from a laser diode,  using the fact that light propagating in the paraxial regime can be modelled with the Schr\"odinger equation.  
The propagation of monochromatic light can be modelled with the Helmholtz equation \cite{Saleh2007}
\begin{equation}
\nabla^{2}U+|\textbf{k}|^{2}U=0,
\end{equation}
where $U$ is a complex amplitude and $\textbf{k}=(k_x,k_y,k_z)$ is the wave vector. 
In the paraxial regime where $|\textbf{k}|\approx k_z$,  this equation can be reduced  to
\begin{equation}
i\frac{\partial}{\partial z}u=\frac{-1}{ 2|\textbf{k}|}\frac{\partial^2}{\partial x^2}u,\label{eq:parax_helmholtz}
\end{equation}
where $U=u\cdot\exp\left(ik_z z\right)$, $u$ is the envelope function of the propagating light,  $z$ is the longitudinal position, and the $y$ coordinate is factored out through a separation of variables.  
Equation~(\ref{eq:parax_helmholtz}) has the form of the one-dimensional Schr\"odinger equation (\ref{eq:SEqnx}) for a free particle. 
Defining an effective mass through $|\textbf{k}|=mc/\hbar$, the correspondence of variables between optical and massive particle regimes is summarized in Table~\ref{table:var_correspondence}. 
Note that we use the transverse angle $\theta=k_x/|\textbf{k}|$, equivalent to a normalized momentum, when plotting results.

\begin{table}[h]
\centering
\begin{tabular}{|c|c|}
\hline
Paraxial Light & Particle in 1D\\ 

[plotted units] & normalized units\\ \hline\hline
Transverse position: $x$ [mm]& Position: $x$ \\ \hline
Longitudinal position: $z$ [m]& Time: $tc$\\ \hline
Transverse angle: $\theta=\frac{k_x}{\left|\textbf{k}\right|}$ [rad] & Momentum: $\frac{p}{mc}$\\ \hline
\end{tabular}
\caption{Variable correspondence between paraxial light and a particle in 1D. Note that in the derivations we use $k_x$ and $|\textbf{k}|$ while the normalized momentum  $\theta$ is used in the plots.} 
\label{table:var_correspondence}
\end{table}

\begin{figure}
\includegraphics[width=\columnwidth]{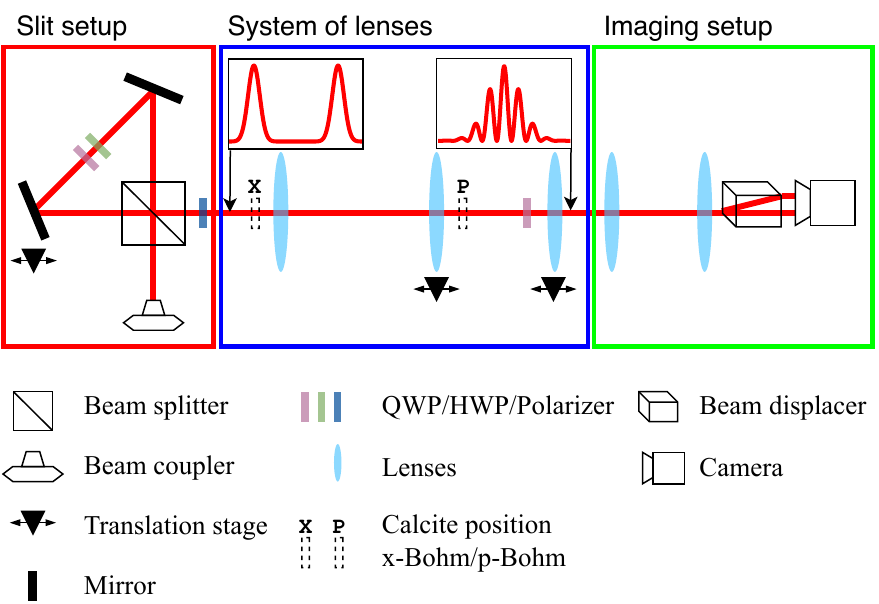}
\caption{Illustration of experimental setup in three parts. 
The slit setup is used to initialize the double-slit experiment.
A diagonally-polarized beam is split into two co-propagating beams using a displaced Sagnac interferometer, where the beam separation (set to 2\,mm throughout the experiment) can be tuned by  translating one of the mirrors. 
The system of lenses is used to simulate propagation of the light along $z$ over a large range (see Fig.~\ref{fig:DistanceCalibration} and Fig.~\ref{fig:Strength_of_weakx} for further details). 
The two plots above the lens system indicate the intensity profiles of the beam before and after the lens transformation. 
A thin piece of calcite is used to weakly couple momentum to polarization (the weak measurement).
The calcite is positioned before the lenses for the weak momentum measurement in the $x$-Bohm experiment and between the lenses for weak position measurement in the $p$-Bohm experiment. 
The imaging setup, consisting of a polarizing beam displacer and a CCD camera, is used to obtain two interference patterns, one for each polarization.
}
\label{fig:fullsetup}
\end{figure}

\begin{figure}
    \centering
    \includegraphics[width=\columnwidth]{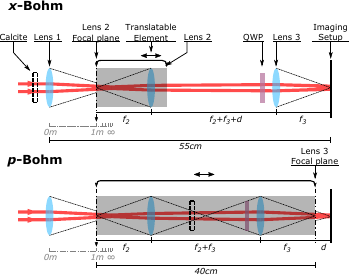}
    \caption{ Illustration of the system of lenses configured to make measurements for $x$-Bohm (top) and $p$-Bohm (bottom) theories. 
    The focal lengths of the lenses are $f_1=$ 15\,cm and $f_2=f_3=$ 10\,cm from left to right, and the total length, from Lens 1 to the imaging setup, is 55\,cm. 
    Lens 1 focuses the beam and remaps the position variable of planes from 0\,m to infinity onto the position variable of planes from 0\,cm to 15\,cm after the lens, with a scaling factor. 
    The grey axis indicates the correspondence between the location of the focus of Lens 2 and the effective propagation distance being imaged by the lens setup. A detailed plot of the propagation distance vs the displacement $d$ of Lens 2 is plotted in Figure \ref{fig:DistanceCalibration}. 
    Top: Lens 2 and Lens 3 map the position variable at the dotted line to the imaging setup (solid line); 
    Bottom: Lens 2 and Lens 3 are set to be 20\,cm away from each other, forming a one-to-one telescope.
  }
    \label{fig:Explain_optics_diagrams_lenses}
\end{figure}

Drawing the analogy between Eq.~\eqref{eq:parax_helmholtz} and the Schr\"odinger equation, we simulate the trajectories of particles in a double-slit experiment by sending 915\,nm laser light through a double-slit apparatus, employing the experimental setup outlined in Figures~\ref{fig:fullsetup} and \ref{fig:Explain_optics_diagrams_lenses}. In the rest of this section we describe the details of this setup, beginning with the gadget used for the weak momentum measurement and the procedure employed in the experimental construction of position trajectories in $x$-Bohm, which closely follow those  outlined earlier~\cite{Kocsis1170,Mahler2016}. We then describe how the same procedure is used to construct momentum trajectories in $x$-Bohm theory, and how the setup is modified for constructing position trajectories in $p$-Bohm theory.

\subsection{Weak momentum measurements}\label{sec:weak_momentum_exp}
A weak measurement is performed by coupling the desired observable to a pointer variable, often a different degree of freedom of the same physical system, followed by a strong pointer variable measurement~\cite{PhysRevLett.66.1107,Dressel2014}. 
Here we use polarization as the pointer. 
Our observable of interest is momentum, which maps to $k_x$ for the light beam. 
We use $\hat{k}_x$ to denote the operator form of $k_x$. 
Specifically, in the position representation $\left\langle x\right| \hat{k}_x \left|x'\right\rangle=-i\partial\delta(x-x')/\partial x$. 
As described below, the shift in polarization will be proportional to the weak value which can then be extracted through a standard polarization measurement. 
We use the notation $\ket{H}$, $\ket{V}$ for horizontal and vertical polarization respectively, $\ket{D} = (\ket{H} + \ket{V})/\sqrt{2}$, $\ket{A} = (\ket{H} - \ket{V})/\sqrt{2}$ for diagonal and anti-diagonal respectively, and $\ket{R}=(\ket{H} +i  \ket{V})/\sqrt{2}$, $\ket{L}=(\ket{H} -i  \ket{V})/\sqrt{2}$ for right- and left-circular polarizations, respectively. 
The pointer is initially set to the diagonal polarization, $\ket{D}$.
Polarization is coupled to the transverse momentum of the light using a thin calcite crystal. 
The interaction can be described by the Hamiltonian
\begin{equation}
    \label{eq:hamiltonian} 
    H_I = \hbar g \frac{\hat k_x}{| \textbf{k} |}  \hat \sigma_z
\end{equation} 
where
\begin{equation}
     \hat \sigma_z = \frac{1}{2} \left(\ket{H}\bra{H}-\ket{V}\bra{V}\right)= \frac{1}{2} \left(\ket{D}\bra{A}+\ket{A}\bra{D}\right),
\end{equation}
and $g$ is the coupling strength. 
If the joint state of the transverse position and polarization before the calcite is $\ket{\Psi} = \ket{\psi} \otimes \ket{D}$, then with a sufficiently weak interaction, i.e., sufficiently small $\zeta k_{x}=gt k_{x}\ll 1 $ over a range of interest for $k_{x}$, the joint state after the calcite is 
\begin{equation}
    \ket{\Psi'} \approx \ket{\psi}\otimes\ket{D} - i\frac{\zeta}{2} \frac{\hat k_x}{|\bf{k}|} \ket{\psi}\otimes \ket{A}.
\end{equation}

The interaction is followed by a  projective $x$ measurement, post-selected on the result, $x_f$.  The state of the pointer following this post-selection is  
\begin{equation}\label{eq:weak_value_rotations_exp_phases}
    \braket{x_f}{\Psi'} \approx \psi(x_f) \left ( e^{-i \frac{\zeta}{2  |\bf{k}  |} \expec{\hat k_x}_w } \ket {H} +  e^{i \frac{\zeta}{2 |\bf{k}|} \expec{\hat k_x}_w } \ket {V} \right ),
\end{equation}
where
\begin{equation}
    \expec{\hat k_x}_w = \frac{\bra{x_f} \hat k_x \ket{\psi}}{\braket{x_f}{\psi}}
\end{equation}
is the weak value, in general a complex number.
The real part of the weak value  shows up as a phase shift between the $H$ and $V$ polarizations and can be extracted by the projective measurement $ \hat \sigma_y  = \frac{1}{2} \left (  \ket{R}\bra{R} - \ket{L}\bra{L} \right )$. 
This corresponds to making measurements of the right- and left-circular intensities, resulting in 
\begin{equation} \label{eq:Sy}
    \expec{ \hat \sigma_y } = \sin \left( \frac{\zeta}{|\bf k |} \text{Re} \left ( \expec{\hat k_x}_w \right ) \right).
\end{equation}

\subsection{Position trajectories in $x$-Bohm theory}
\label{sec:exp_xbohm_pos_traj}
The double-slit pattern is generated by separating a Gaussian beam ($1/e^2$ diameter of 0.55\,mm) into two, using a horizontally-displaced Sagnac interferometer (slit setup in Figure~\ref{fig:fullsetup}), which gives an effective slit separation of 2\,mm.
The light is then diagonally polarized and sent through a thin calcite crystal (0.2\,mm, cut at 45 degrees) to weakly couple the transverse momentum of the light to polarization via a birefringent phase shift (see Sec.~\ref{sec:weak_momentum_exp} above). 
Importantly, the interaction Hamiltonian \eqref{eq:hamiltonian} commutes with the Hamiltonian for free propagation. This implies that the calcite crystal can remain fixed at a single $z$ position before the lens system independent of the plane of interest.

Next, the co-propagating beams traverse a system of three lenses (Fig.~\ref{fig:fullsetup}, middle pane), labelled Lens 1, 2, 3 with respective focal lengths 15\,cm, 10\,cm, 10\,cm (see Figure~\ref{fig:Explain_optics_diagrams_lenses}). 
By translating Lens 2 along the $z$-axis, we simulate different propagation distances for the light, resulting in effective distances ranging from $0.66$\,m to $3.5$\,m.
In other words, the three-lens system maps what would have been the transverse position of the light beam propagating in free space onto the transverse position at the end of the lens system\footnote{With our experimental parameters, $\lambda = 915$\,nm, slit separation $s= 2$\,mm, and slit width $w = 0.55$\,mm, we expect the near-to-far field transition to occur at $\frac{s}{2} / (\lambda / (\pi w/2))  = 0.77$\,m.}. The calibration of the lens system is discussed in Appendix~\ref{sec:calibrationsec}.

Finally, the co-propagating beams enter the imaging setup (Fig.~\ref{fig:fullsetup}, right pane), where the resulting intensity patterns at the end of the lens system were measured on a CCD camera. 
In addition to the intensity of the interference pattern, the polarization is measured by a quarter wave plate and a polarization beam displacer that effectively separates the left- and right-circularly polarized light in the vertical direction.
Since the interference occurs along the horizontal transverse axis\footnote{This is the axis that is horizontal and perpendicular to the axis of propagation.}, the interference patterns for the left- and right-circular polarizations can be measured independently.  
The intensity patterns of the two polarizations ($|u|^2$ in Eq.~(\ref{eq:parax_helmholtz})), given by $I_R$ and $I_L$, differ by an amount directly related to the real part of the weak value of transverse momentum, which in the limit of an infinitely weak measurement can be extracted as 
\begin{equation}
\frac{ \text{Re} \left ( \expec{\hat{k_x}}_w \right )}{\left | \textbf{k} \right |} = \frac{1}{\zeta} \left[\sin^{-1} \left ( \frac{I_R - I_L}{I_R + I_L} \right )-\phi_0\right],
\label{eq:birefringent}
\end{equation}
where the $\sin^{-1}$ term comes from Eq.~(\ref{eq:Sy}) and $\phi_0$ is a momentum-independent phase shift acquired in the calcite crystal, set by tilting the calcite; $\zeta=134.49\pm0.13$ is the coupling strength\footnote{The quantity $\zeta$ corresponds to rotation imparted to the polarization of light per transverse angle of the light, and is hence dimensionless.} which depends on the length of the calcite. 
The calibration of $\zeta$ is discussed in Appendix~\ref{sec:calibrationsec}.
Examples of measured intensity patterns for near- and far-field propagation distances are shown in the top row of Figure~\ref{fig:Selected-planes}, while measured values of the momentum, in the same two planes, are shown in the third row of Figure~\ref{fig:Selected-planes}.
By performing this measurement for each $z$-plane, we extracted ensemble-average values of the transverse momentum as a function of position, from which we construct particle trajectories.
Experimentally constructed $x$-Bohm position trajectories are shown in the top row of Figure~\ref{fig:Traj_overlaid}, with theoretically calculated trajectories shown in the bottom row.
We will discuss all experimental results in greater detail in Section~\ref{sec:results}.

\subsection{Momentum trajectories in $x$-Bohm theory}
To construct the momentum trajectories in $x$-Bohm we follow the procedure outlined in Sec.~\ref{sec:BohmTraj_nonstandard}, where again we use Eq.~\eqref{eq:pinx}.
In our case the momentum is proportional to the velocity (see Eq.~(\ref{eq:pmv})), which implies that the measurement performed to construct the $x$ trajectories in Section~\ref{sec:exp_xbohm_pos_traj} suffices for constructing the momentum trajectories. 
The resulting trajectories are shown in the first row of Figure~\ref{fig:PTraj_overlaid}.

Note that proportionality between velocity and momentum is only valid when the potential term in the Hamiltonian is independent of $p$. 
In cases where the potential has  $p$ and/or $p^2$ terms, Eq.~(\ref{eq:Bohm_velocity_operational}) is no longer valid, while Eq.~(\ref{eq:pinx}) remains valid generically. 

\subsection{Momentum trajectories in $p$-Bohm theory}
As described in Sec.~\ref{sec:pbohm_intro}, the conservation of momentum for a free particle implies that the $p$-Bohm momentum trajectories follow lines of constant $p$. 
There is no need to construct these trajectories experimentally; however, the relative probabilities (or density of trajectories) can be measured by making a strong $p$ measurement. 
In practice this is accomplished by strong $x$ measurements in the far field,  using the fact that momentum maps to position at infinity (see Figure~\ref{fig:Strength_of_weakx} in Appendix~\ref{sec:calibrationsec}). 
The same setup is used to calibrate the coupling strength of the weak measurement (again, see Appendix~\ref{sec:calibrationsec}).  
The resulting trajectories are shown in the second row of Figure~\ref{fig:PTraj_overlaid}.

We emphasize that  $p$-Bohm theory in three dimensions is not unique, and that different theories lead to different expressions ${\bf{v}}_p({\bf{p}},t)$ for the momentum-space velocity \cite{Struyve_2008}. However, in one dimension the continuity constraint \eqref{eq:momentum_continuity} essentially identifies \eqref{eq:momentum current} as the current density in momentum-space, which in the limit of a free particle leads via Eq.~\eqref{eq:momentum_velocity}  to the conservation of the $p$-Bohm momentum. The results in Fig.~\ref{fig:PTraj_overlaid} are therefore free of any ambiguity that would affect $p$-Bohm theories in higher dimensions. 

\subsection{Position trajectories in $p$-Bohm theory}
\begin{figure*}
\centering
\includegraphics[width=0.9\linewidth]{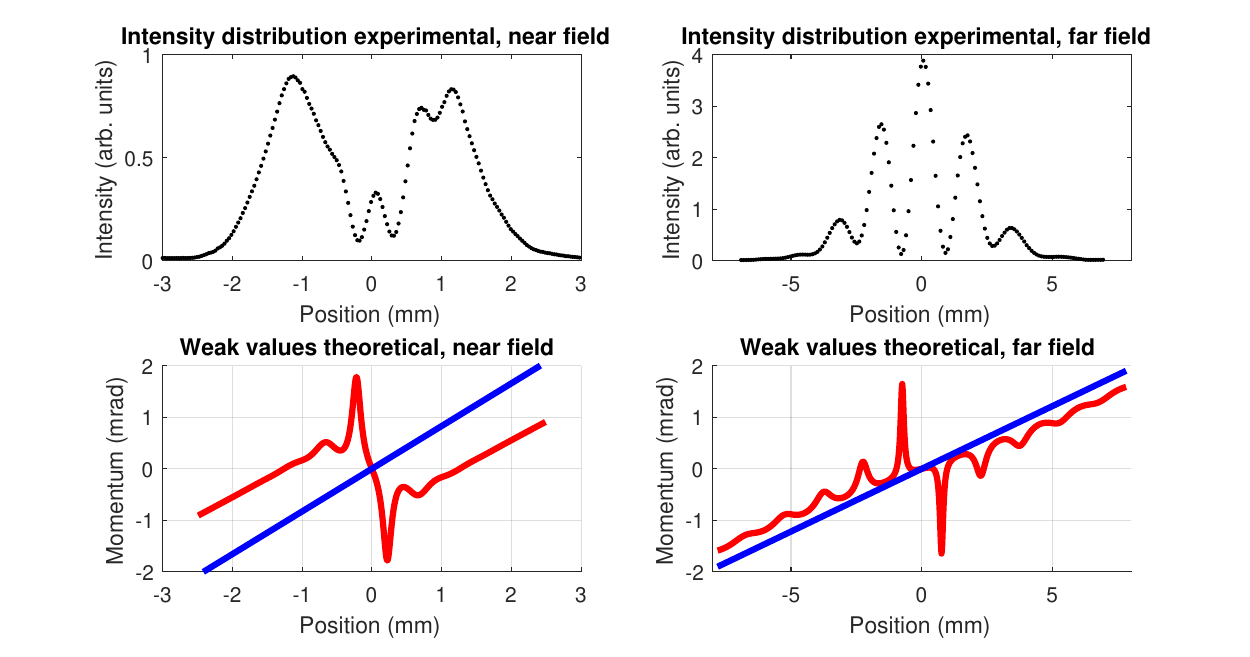}
\includegraphics[width=0.9\linewidth]{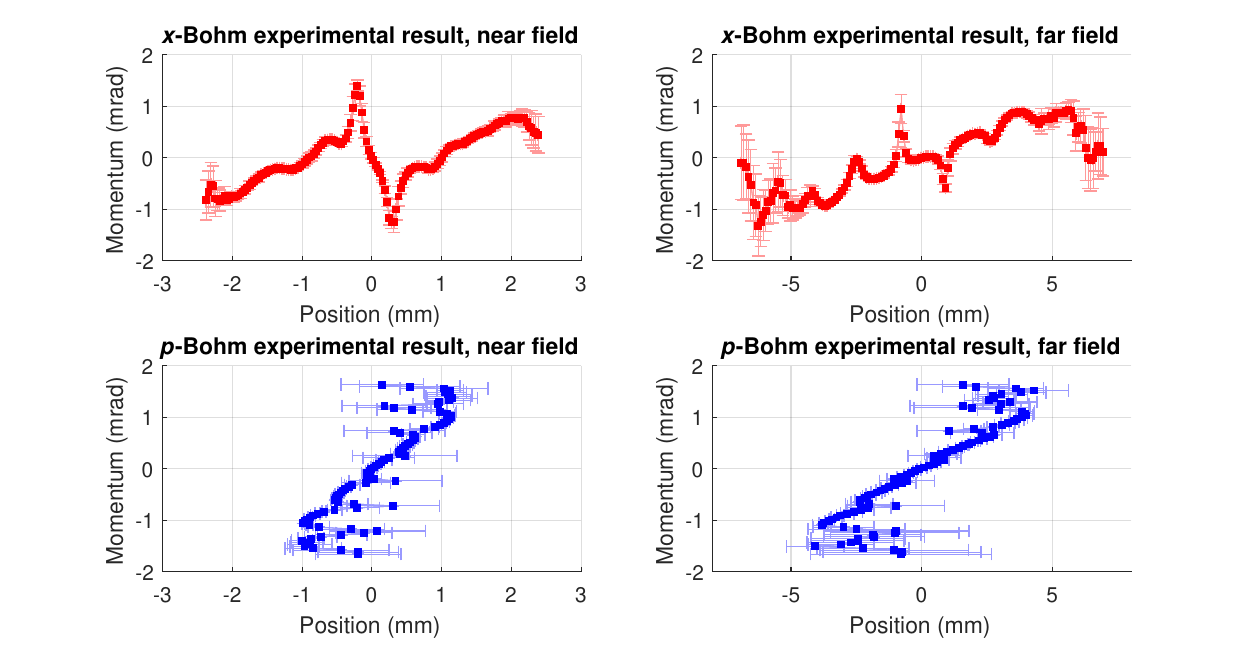}
\caption{ Near field and far field snapshots. Near field (left column) data were taken at an effective $0.70\,$m from the slits, and far field (right column) taken at  $3.5$\,m.
The transition from near- to far-field occurs around 0.77 m, given by the ratio of the slit separation to the full divergence angle of the beams.
The intensity profiles displayed in the first row, i.e., the position distributions, are the sum of left- and right-circular intensities and show typical two-slit interference patterns. 
The asymmetry in the intensity distributions results from a slight difference in intensities between the two slits. 
Note that the far-field intensity profile differs slightly from the momentum distribution, which has a higher interference visibility.
Numerical simulations of the Bohmian position-momentum profiles for $x$-Bohm (red) and $p$-Bohm (blue) theories in the second row show qualitative differences between the predictions of the two theories, in particular in the near field. 
The results of the two theories are expected to converge at infinity.
Third and fourth rows show the measured position-momentum profiles for $x$-Bohm and $p$-Bohm theories respectively.
A qualitative difference between the predictions of the two theories can be seen, notably the presence of strong peaks in the near-field data for $x$-Bohm and not for $p$-Bohm.
Error bars correspond to the standard deviation between weak values given by different calcite tilt angles.
Error bars are larger in areas where the overall intensity of the interference is small (see  Section~\ref{sec:single_time}).
Note that in the $p$-Bohm experiment post-selection is always at infinity so that the minima do not correspond to those in the intensity profiles in the first row. The slope (near field: $1.03m\:rad^{-1}\pm 0.01m\:rad^{-1}$; far field: $3.70m\:rad^{-1}\pm 0.09m\:rad^{-1}$) of the $p$-Bohm experimental weak value given by linear fitting data within the momentum range of $[-1,1]mrad$ is slightly different from the theoretical weak value (near field: $1.21m\:rad^{-1}$; far field: $4.11m\:rad^{-1}$). It is important to note that the fitted value of the slope is biased towards zero due to data points corresponding to low post-selection probability being biased towards zero. Additionally, there were systematic uncertainties not reflected in the error bars associated with the calibration of the effective propagation distance.
 \label{fig:Selected-planes}}
\end{figure*}

Position is a non-primary variable in $p$-Bohm theory, with the procedure of constructing its trajectory given in Sec.~\ref{sec:BohmTraj_nonstandard} using Eq.~\eqref{eq:xinp}. 
This amounts to making weak position measurements followed by a post-selection on momentum. 
To perform the weak measurement,  Lens 2 and Lens 3 are kept at a fixed distance from one another, making a one-to-one telescope, and are translated together (Figure~\ref{fig:Explain_optics_diagrams_lenses}).  
In this way, the transverse momentum between Lens 2 and Lens 3 corresponds to transverse position in a fixed propagation plane.
As such, the calcite crystal is placed in between Lens 2 and Lens 3 to perform a weak position measurement. 
Additionally, the one-to-one telescope relates the light one focal length before Lens 2 to the light one focal length after Lens 3 by an identity transformation.
This effectively places the far field of the interfering beams onto the imaging setup, causing it to perform a strong momentum measurement.
To read out the weak measurement, the quarter wave plate and polarization beam displacer, once again, are used to separate the left- and right-circularly polarized component of the beam in the vertical direction and, with procedures similar to those in Section~\ref{sec:exp_xbohm_pos_traj}, we can extract the weak position value post-selected on momentum.
The fourth row of Figure~\ref{fig:Selected-planes} shows results of the corresponding weak measurements in near- and far-field planes.
Position trajectories, with momentum as the primary ontological variable, are constructed in the same manner as before. 
Constructed trajectories are shown in the second row of Figure~\ref{fig:Traj_overlaid}.

\section{Comparison of $x$-Bohm and $p$-Bohm trajectories}
\label{sec:results}

We now consider the $x$-Bohm and $p$-Bohm particle trajectories in detail.
The trajectories are constructed by interpolating data points taken at discrete $z$-planes ranging from an effective distance of $0.66$\,m to $3.5$\,m after the slits. 
The results presented and discussed below show the qualitatively different behavior of the trajectories in the two theories, especially in the near-field. 
The experimental results are in good agreement with the theoretical predictions, and illustrate the dependence of the ontological description on the choice of the primary ontological variable.

\subsection{Single-time position-momentum snapshots}\label{sec:single_time}
For a given $z$-plane, which corresponds to an instant in time, data were taken by fixing the lenses and post-selecting on the primary ontological variable producing a complete description of the functions $p\left(x,t\right)$ and $x\left(p,t\right)$ for the $x$-Bohm and $p$-Bohm theories respectively. 
Results for two of these instants of time, one in the near-field and one in far-field, are presented in Figure~\ref{fig:Selected-planes} and compared with theoretical predictions. 
To illustrate the difference between the two theories we begin with a numerically simulated plot (Figure \ref{fig:Selected-planes}, second row), where we overlay two ontological momentum-position snapshots, based on Eq.~\eqref{eq:Holland}. Experimental results are shown in the third and fourth rows of Figure~\ref{fig:Selected-planes}. 

In $x$-Bohm, peaks in the momentum $p\left(x,t\right)$ appear when a particle approaches a minima in the double-slit interference pattern (i.e. the minima in the top row of Figure~\ref{fig:Selected-planes}). 
These peaks get progressively narrower, with width approaching zero, as the measurement is taken further into the far field. 
The asymptotic large $x$ behavior of the function $p\left(x\right)$ corresponds to
\begin{equation}\label{eq:asymp_xBohm}
    p/m=\frac{x-\text{sgn}(x) w/2}{t},
\end{equation}
with $t$ being propagation time and $w=2$\,mm being the slit separation. This can be roughly interpreted as the consequence of the guiding wave $\psi(x,t)$ at the near field having two distinguishable parts with a small overlap so that particles away from the overlap are effectively guided by one or the other, leading to behaviour similar to what one would observe if only one slit were open. 

In $p$-Bohm theory, we expect a linear relation $x\left(p\right)=p\cdot t/m$. 
It is important to note that, experimentally, data with momentum post-selection always projects the far field onto the imaging setup. Similarly, the guiding wave  $\tilde \psi(p,t)$  has the form of the far-field interference pattern. 


Due to the nature of our measurement, the weak value of the variable of interest is very sensitive to background noise when the post-selection probability is small. 
When background light and other systematic errors dominate the measured signal, the probability of registering a measurement in the left- and right-circular polarization basis becomes roughly equal.
As a consequence, and by referring to Eq.~\eqref{eq:birefringent}, one can see that the weak value tends to the incorrect result of $-\zeta^{-1}\phi_0$ near the minima of the interference patterns. 
As mentioned in Section~\ref{sec:exp_xbohm_pos_traj}, the value of $\phi_0$ in our experiment is controlled by the horizontal tilt angle of the calcite crystal.
In an idealized noiseless measurement, the value of $\phi_0$ exists purely as a calibration parameter of the weak measurement (see Appendix~\ref{sec:calibrationsec}) and does not affect the measured weak value. 
However, with some amount of noise present in the measurements this is not the case. 
To account for this imperfection, we measured weak values using various calcite tilts.
The final weak value at each time is tabulated by averaging measurement results with different calcite tilts.
Error bars in the  third and fourth row of Figure~\ref{fig:Selected-planes}  correspond to the standard deviation of the measurement given a set of values for $\phi_0$. The effects of the calcite tilt are particularly pronounced in $p$-Bohm experiments, where the measurements at the minima go very close to zero and the standard deviation between measurement results increases significantly.

\begin{figure*}
\includegraphics[width=\linewidth]{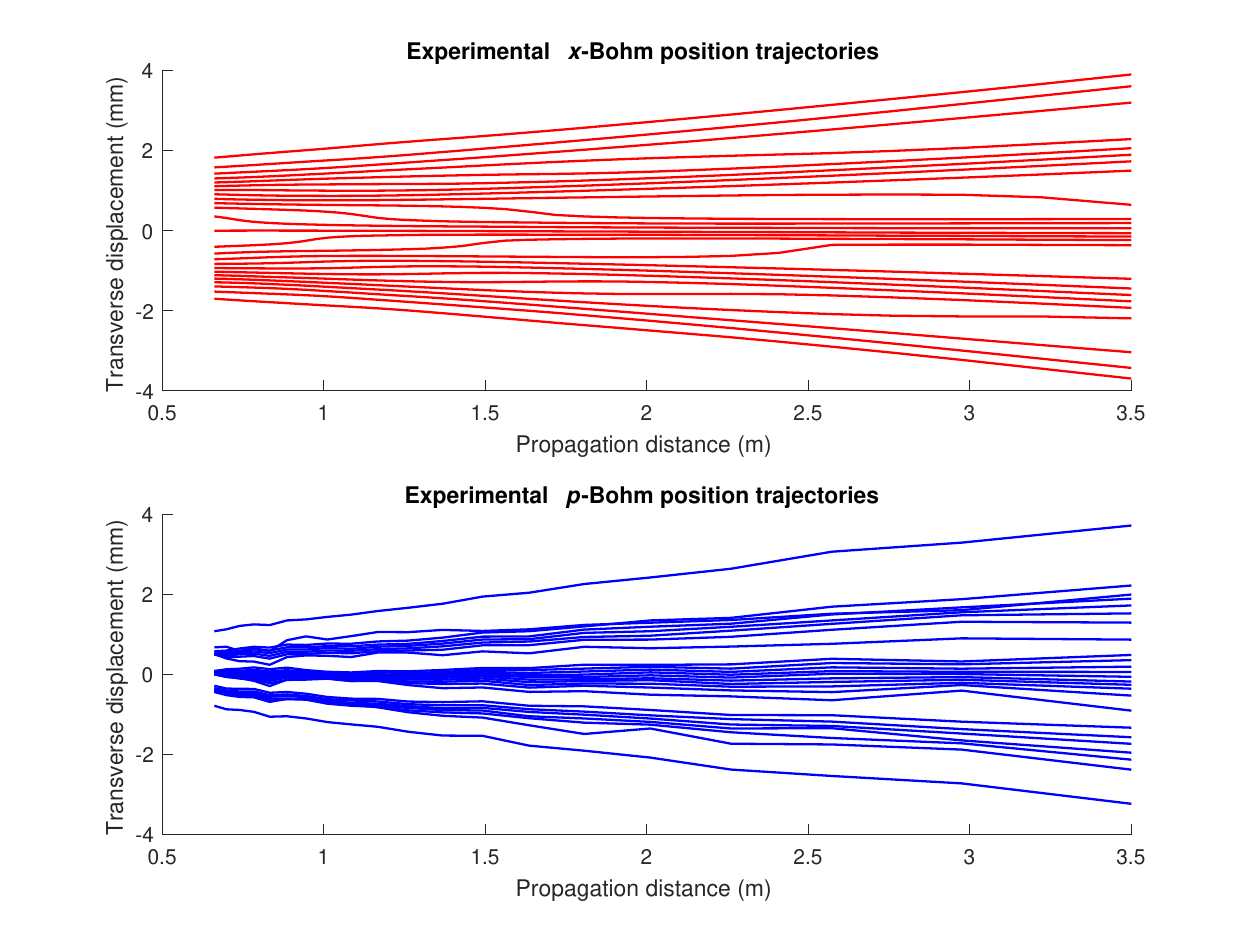}
\includegraphics[width=\linewidth]{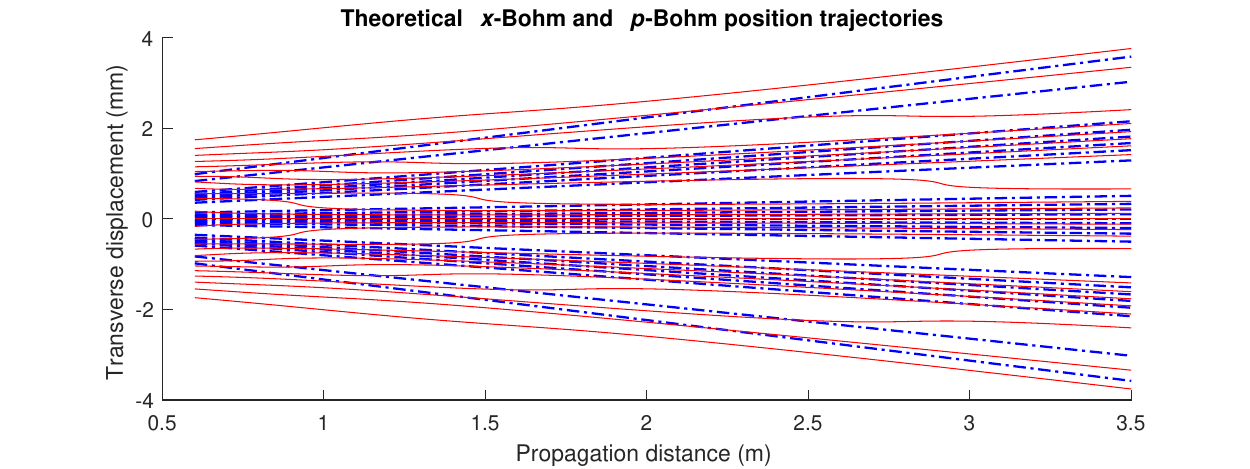}
\caption{Constructed position trajectories based on $x$-Bohm (red) and $p$-Bohm (blue) theories.
Top (middle) plot corresponds to $x$-Bohm ($p$-Bohm) position trajectories constructed experimentally.
Bottom plot corresponds to numerical simulation of both $x$-Bohm (red solid line) and $p$-Bohm (blue dotted line) position trajectories overlaid. 
$x$-Bohm trajectories originate from the location of the two slits, while $p$-Bohm trajectories originate from mid-point between the two slits. 
In the far field, trajectories from $x$-Bohm and $p$-Bohm theory converge to the same values as expected.
The $x$-Bohm trajectories are also plotted in Figure \ref{fig:XBohm_Selected_Traj} (top) with one position trajectory highlighted.
\label{fig:Traj_overlaid}}
\end{figure*}

\begin{figure*}
\includegraphics[width=\linewidth]{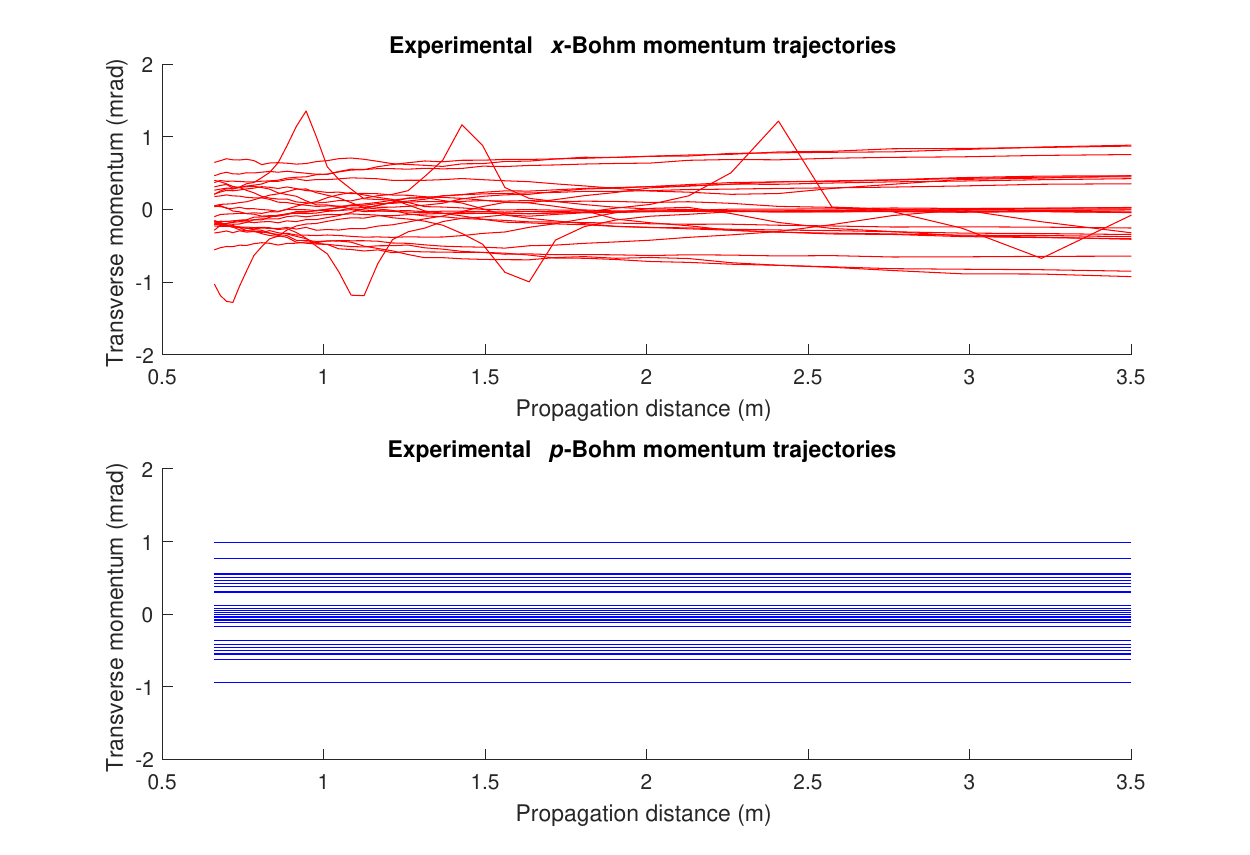}
\includegraphics[width=\linewidth]{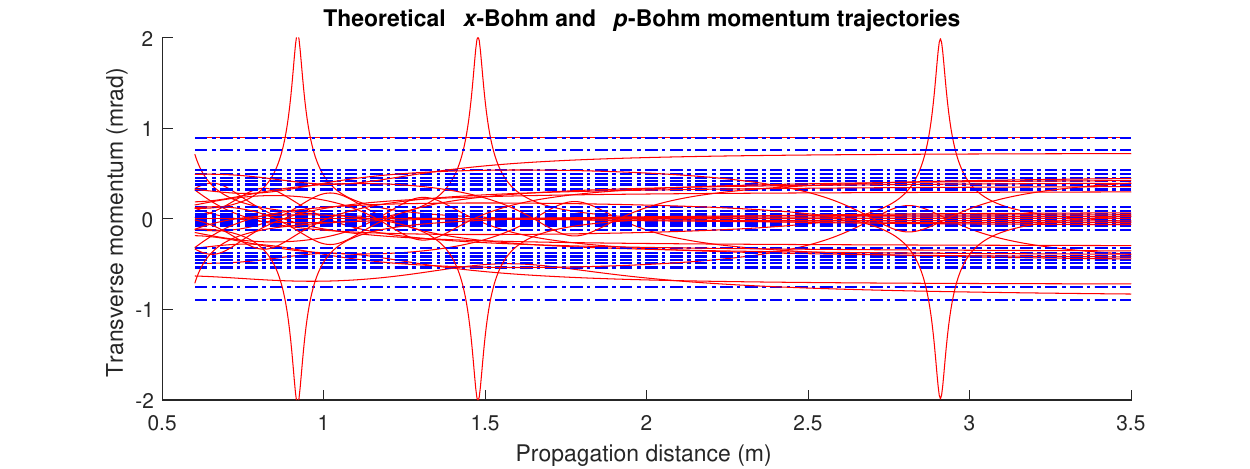}
\caption{Constructed momentum trajectories based on $x$-Bohm (red) and $p$-Bohm (blue) theories.
Top(middle) plot corresponds to $x$-Bohm($p$-Bohm) momentum trajectories constructed  experimentally.
Bottom plot corresponds to numerical simulation of both $x$-Bohm (red solid line) and $p$-Bohm (blue solid line) momentum trajectories overlaid. 
In the far fields both sets of trajectories bunch in the manner reflective of the far field interference pattern. The transverse momentum along the momentum trajectories for $x$-Bohm are equal to the first derivative of the position along the position trajectories under $x$-Bohm, whereas the momentum trajectories for $p$-Bohm are flat lines due to the conservation of momentum. 
Peaks in the $x$-Bohm trajectories corresponds to the crossing of the particle over an interference minimum, and their time of occurrence is highly sensitive to the initial conditions of the primary ontological variable, causing inconsistencies between the numerically simulated and experimentally constructed trajectories. 
The design of our experiment presumes conservation of momentum. As a result, the momentum trajectories given by $p$-Bohm cannot be anything other than flat, reflecting momentum conservation as a built in assumption in our experiment.
The $x$-Bohm trajectories are also plotted in Figure \ref{fig:XBohm_Selected_Traj} (bottom) with one momentum trajectory highlighted.   
\label{fig:PTraj_overlaid}}
\end{figure*}

\begin{figure}
\center{\includegraphics[width=1.08\linewidth]{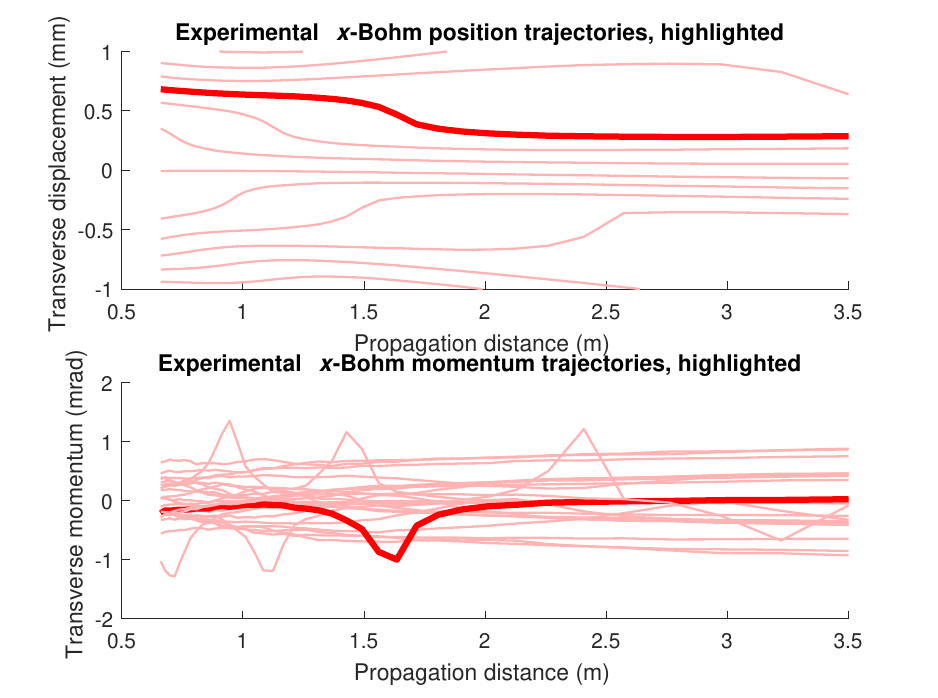}}
\caption{$x$-Bohm experimental trajectories for position (top) and momentum (bottom) with a single initial setting highlighted for clarity. 
Note that the highlighted, non-primary variable, trajectory of momentum (bottom) corresponds to the highlighted ontological trajectory of position (top). 
Peaks in the momentum trajectories correspond to the ontological position trajectories crossing a minimum in the two-slit interference pattern. 
\label{fig:XBohm_Selected_Traj}}
\end{figure}


\subsection{Constructing Trajectories}\label{sec:traj_results}

We construct a set of trajectories for both position and momentum in both $x$-Bohm and $p$-Bohm theory, chosen so that the density of the selected trajectories in the primary ontological space corresponds to particle distribution probabilities.  
This is possible due to the fact that the velocities are defined through the probability current (see Sec. \ref{sec:BohmTheory}). However, there is no a priori reason to expect this feature to be preserved in the non-primary space, and indeed we will see that it is not.
Similarly, the trajectories in the primary space cannot cross, since given a wave function, the velocity is uniquely defined by the value of the primary ontological variable. 
As we will see, the $p(x,t)$ trajectories in $x$-Bohm do cross.    
The trajectories for $x$-Bohm and $p$-Bohm experiments  are shown in  Figures \ref{fig:Traj_overlaid} and \ref{fig:PTraj_overlaid} respectively along with theoretical trajectories derived from a numerical simulation.

\subsubsection{Position Trajectories}\label{sec:pos_traj_results}
\label{sec:Pos_Traj_Results}
The $x$-Bohm position trajectories shown in the top row of Figure~\ref{fig:Traj_overlaid} are very similar in nature to those obtained earlier~\cite{Kocsis1170}.
These trajectories originate from one of the two slits and, while providing the signatures of interference for the  probability density, they generally diverge away from $x=0$ while displaying a rapid `acceleration' through each region of destructive interference, where the density becomes low and the ratio of flux to density correspondingly large.
As required by the Bohmian formalism, the trajectories of the primary ontological variable do not cross. 

For $p$-Bohm, the position trajectories (Figure~\ref{fig:Traj_overlaid} middle) originate from a single point in between the two slits and spread out in a manner that preserves momentum, resulting in straight lines given by $x=pt/m$. 
A crossing (or in this case convergence to a point at $t=0$) is possible since these are not the trajectories of the primary ontological variable. 
Imperfect translation of Lens 2 and Lens 3 causes some transverse displacement, leading to a systematic error in the weak value measurement such that the trajectories are displaced by a different amount at each plane. This causes the trajectories to shift in the y-axis of Figure~\ref{fig:Traj_overlaid}, resulting in the experimental weak values of position deviating from simple straight lines.

Apart from the obvious discrepancy between the position trajectories in  $x$-Bohm and $p$-Bohm, the $p$-Bohm position trajectories  exhibit the potentially surprising phenomenon of originating at $x=0$, rather than in either of the slits.
That is, the initial position for all the particles according to p-Bohm theory is a position which has vanishingly low probability according to $x$-Bohm; moreover, a detector placed at that position would never be expected to register a photon.
Note that since the value of position in $p$-Bohm is given by the weak value (\ref{eq:pinx}), it is different from the result of the projective position measurement that is associated with the double slit or a detector. In general, the values of non-primary variables do not need to agree with the results of projective measurements in standard quantum theory. Indeed in the $p$-Bohm picture, the initial state prepared in our experiment sets the position variable to be zero.

\subsubsection{Momentum Trajectories}
\label{sec:Mom_Traj_Results}
Next, we construct particle momentum trajectories, tracking the change of momentum over time (see Figure~\ref{fig:PTraj_overlaid}). 
As the conservation of momentum of light in free space is an assumption used in the alignment, the $p$-Bohm momentum trajectories are constructed from theory as flat lines with a distribution derived from the strong momentum measurement (position in the far field).

The $x$-Bohm momentum trajectory functions are proportional to the time derivative of the position trajectory functions.
The peaks observed in $x$-Bohm momentum trajectories correspond to time intervals when the position trajectories are crossing the minima of the interference pattern (as emphasised in Fig. \ref{fig:XBohm_Selected_Traj}). 
The time instances at which these peaks appear are highly sensitive to the initial conditions of the $x$-Bohm position trajectories, and as such, they do not align with the peak positions in the numerical simulation. These peaks indicate the non-conservation of momentum in $x$-Bohm for individual particle trajectories, which is analogous to the issue of $p$-Bohm position trajectories originating in between the two slits discussed in section \ref{sec:pos_traj_results}.

To further explain the behaviour at the peaks, we highlight a single trajectory line in Figure~\ref{fig:XBohm_Selected_Traj}, where the top and bottom plots correspond to a position and momentum trajectory in $x$-Bohm for the same initial conditions. 
A peak in the momentum trajectory directly corresponds to the portion of the position trajectory where the particle crosses a minimum in the double-slit interference pattern.

\section{Discussion}
\label{sec:discussion}
\begin{quotation}
When everyone is somebody, then no one’s anybody.\\
 \textit{W.S. Gilbert, The Gondoliers}\\
\end{quotation}
The lack of an ontological interpretation has been criticised as a serious drawback of quantum theory since its early days \cite{QuantumTheoryCrossroads}. Part of the deeper understanding that an ontological interpretation would provide would be a visualization of the underlying quantum dynamics. 
Apart from any philosophical considerations, such visualizations are arguably essential for developing the intuition
necessary for scientific development. At the same time, incorrect visualizations (such as those involving the aether in electrodynamics)
can lead us astray.
Bohm's interpretation, with its deterministic particle trajectories, presents an attractive visual picture of quantum
dynamics at the cost of some non-trivial assumptions.
Among these is an assumption of the role of position and the coordinate representation
of the wave function as the fundamental variables that determine the dynamics. 
Indeed, in contrast to classical physics, where both the initial position and momentum are necessary for predicting the dynamics
that follows, in an $x$-Bohm theory the initial conditions are just the initial position of the particle, together with the initial wave
function. 
It is therefore tempting to view the identification of the asymmetry as a profound discovery, suggesting that position is indeed more important than other variables.
One could even hope that the realization that position plays a special role would lead to new experimental predictions. 
However, as emphasized in this work, the specific choice of position is not unique, leaving us with an infinite number of possible primary ontological variables -- each allegedly more fundamental than all the others -- or, as Gilbert's line above implies, with none. 

Our main results show that a variation of Bohm's theory ($p$-Bohm), one in which the primary ontological variable is momentum, leads to very different phase-space dynamics. 
In this theory the fundamental trajectories are paths through momentum-space, and the equations of motion take a form which is closer to that of Newton’s laws, with a first time-derivative for momentum. 
If the underlying phase-space dynamics of both theories were the same, one might say that the symmetry between position and momentum had been restored, removing a non-trivial assumption from Bohm's theory. 
The pictures are, however, very different (see Figures \ref{fig:Selected-planes}, \ref{fig:Traj_overlaid} and \ref{fig:PTraj_overlaid}) and so, the asymmetry in a Bohm-like theory is confirmed but ambiguous. 
One is left to wonder which of the two theories, or indeed of the infinite intermediate theories with other ontology, is correct, and possibly more importantly what is the primary ontological variable.

A striking example of this conundrum arises when we consider a harmonic oscillator, where the Hamiltonian operator $\hat H$ can be written in terms of the usual raising and lowering operators $\hat a^{\dagger}$ and $\hat a$, 
\begin{align}
 & \hat H=\hbar\omega \left (\hat a^{\dagger} \hat a + \frac{1}{2} \right ).\label{eq:harmonic oscillator}
\end{align}
We can also write 
\begin{align}
 & \hat H=\frac{1}{2} \hat p_{\theta}^{2}+\frac{1}{2}\omega^{2} \hat x_{\theta}^{2},
\end{align}
for any real $\theta$, where the operators $\hat x_{\theta}$ and $\hat p_{\theta}$
are defined as 
\begin{align}
 & \hat x_{\theta}=\sqrt{\frac{\hbar}{2\omega}}\left( \hat a^{\dagger}e^{i\theta}+ \hat a e^{-i\theta}\right),\\
 & \hat p_{\theta}=i\sqrt{\frac{\hbar\omega}{2}}\left(\hat a^{\dagger}e^{i\theta}- \hat a e^{-i\theta}\right).
\end{align}
Since  $\left[\hat x_{\theta}, \hat p_{\theta}\right]=i\hbar$,
we can construct what might be called a $\theta-$Bohm theory by taking $\hat x_{\theta}$ to be the ``position operator'' for the particular $\theta$ chosen; in this representation the Schr\"odinger equation is 
\begin{align}
 & i\hbar\frac{\partial}{\partial t}\psi(x_{\theta},t)=-\frac{\hbar^{2}}{2}\frac{\partial^{2}\psi(x_{\theta},t)}{\partial x_{\theta}^{2}}+\frac{1}{2}\omega^{2}x_{\theta}^{2}\psi(x_{\theta},t),
\end{align}
and following the usual Bohmian procedure and taking $\psi(x_{\theta},t)=R_{\theta}(x_{\theta},t)\exp(iS_{\theta}(x_{\theta},t)/\hbar)$, with $R_{\theta}(x_{\theta},t)$ and $S_{\theta}(x_{\theta},t)$ both real, the guidance equation is
\begin{align}
 & v_{\theta}(x_{\theta},t)=\frac{1}{2}\frac{\partial S_{\theta}(x_{\theta},t)}{\partial x_{\theta}}.
\end{align}
At least following Holland's suggestion~\cite{Holland_book}, this would be taken as the value of the non-primary variable associated with $\hat p_{\theta}$ (compare Eq.~\eqref{eq:Holland}).
Here for each $\theta$ a different physical picture emerges, and weak $p_{\theta}$ measurements followed by strong $x_{\theta}$ measurements would allow the construction of trajectories for each $\theta$-Bohm theory, yielding entirely different visualizations.

The significance of this is apparent if one considers the Hamiltonian (\ref{eq:harmonic oscillator}) to describe a mode of the radiation field associated with a standing wave. 
Then in a standard treatment~\cite{GerryKnight} the operators $\hat x_{0}$ and $\hat p_{0}$ are (within factors) associated with the electric and magnetic fields respectively. 
Thus in the $0$-Bohm theory the ground state (or indeed any energy eigenstate) would be associated with an ensemble of different values of the electric field but, following Holland's suggestion, the magnetic field would vanish in each member of the ensemble. 
On the other hand, in the $\pi/2$-Bohm theory it would be $\hat p_{\pi/2}$ that would correspond to the electric field and $\hat x_{\pi/2}$ with the magnetic field, and so a description of the ground state (or indeed any energy eigenstate) in the $\pi/2$-Bohm theory would be associated with an ensemble of different values of the magnetic field, but with the electric field vanishing in each member of the ensemble. 
Yet other descriptions would arise for the ground state for other values of $\theta$. 
Considering more general quantum states of the radiation field, weak measurements associated with one field quadrature followed by strong measurements associated with the complementary quadrature would allow for the formal construction of very different sets of trajectories.

What would be the physical motivation for granting reality to one set of trajectories or the other?  It has been suggested that the question can be settled in favour of the usual position variable if one considers a potential with an $x$ dependence more than quadratic, first by Wiseman by examining the relation between weak measurements and the continuity equation~\cite{Wiseman2007}, and then by Gambetta and Wiseman by examining the convergence of the continuum limit of Bell’s modal dynamics and $x$-Bohm~\cite{Gambetta2004}.  Indeed if one were to describe the potential involved in preparing the double slit state in our experiment, it would have terms that are higher than quadratic.
Unfortunately the measurement of trajectories in such theories remains experimentally challenging even with the simplification of a photonic simulation. 
We expect that continued work in this direction, ideally experiments involving massive particles, would lead to results that shed further light on the question. 
For states of the radiation field, weak and strong measurements of field quadratures would extend the discussion of the kind of issues raised here to Bohmian descriptions of field theories.

\begin{acknowledgements}  
  This work was supported by NSERC and the Fetzer Franklin Fund of the John E. Fetzer Memorial Trust. A.M.S. is a fellow of CIFAR.  We thank David Schmid for useful discussions.  
\end{acknowledgements}

\bibliography{Bohm_paper_citations}

\onecolumngrid

\appendix
\section{ABCD Matrix Transformation}
To calculate the effective imaging plane, we employ the ray transfer matrix analysis on our lens system \cite{Saleh2007}. The propagation and thin lens matrix is given by
\begin{subequations}
\begin{eqnarray}
 M_{\rm{prop}}\left(d\right)=\left(\begin{matrix}1&d\\0&1\end{matrix}\right),\label{eq:ABCD_prop} \\
 M_{\rm{lens}}\left(f\right)=\left(\begin{matrix}1&0\\-\frac{1}{f}&1\end{matrix}\right).\label{eq:ABCD_lens}
\end{eqnarray}
\end{subequations}
To see the effective imaging plane at some distance $y$ from the slits after the Lens 1, we analyze the ray matrix  of light being back propagated for a distance of $y$ and then forward propagated through a lens and some distance $f_1-d$. 
This results in a ray matrix of 
\begin{equation}
    \left(
    \begin{array}{cc}
     1-\frac{f_1-d}{f_1} & -y \left(1-\frac{f_1-d}{f_1}\right)-d+f_1 \\
     -\frac{1}{f_1} & \frac{y}{f_1}+1 \\
    \end{array}
    \right).
\end{equation}
For the position distribution of the plane after the lens to be equivalent to the effective imaging plane, we require
\begin{equation}
    d=\frac{f_1}{1+y/f_1}.\label{eq:effective_plane_condition}
\end{equation}
This results in the transformation matrix 
\begin{equation}
    \left(
    \begin{array}{cc}
     \frac{f_1}{f_1+y} & 0 \\
     -\frac{1}{f_1} & \frac{f_1+y}{f_1} \\
    \end{array}
    \right),
\end{equation}
which corresponds to a transformation that relates the position of the beam as a function of the position of the beam before the transformation only and does not depend on the momentum of the beam before the transformation. The resulting transformation also results in a scaling of $f_1/(f_1+y)$ in the position variable from the effective image plane to the plane at distance $f_1-d$ after Lens 1. 
Placing the focus of Lens 2 at a distance of $f_1-d$ after Lens 1 results in a transformation matrix of 
\begin{equation}
    \left(
\begin{array}{cc}
     \frac{f_1 (f_2-d_2)}{f_2 (f_1+y)}-\frac{f_2}{f_1} & \frac{f_2 (f_1+y)}{f_1} \\
     -\frac{f_1}{f_1 f_2+f_2 y} & 0 \\
    \end{array}
\right)
\end{equation}
where $d_2$ is the distance after the second lens. 
As one can conclude from inspecting this matrix, the momentum of the light after the second lens is independent of the momentum at the effective imaging plane and proportional to its position distribution. 
Placing a calcite to perform a weak momentum measurement after Lens 2 thus results in the weak position measurement at the effective image plane.

Lens 3 controls whether position or momentum of the effective imaging plane is projected onto the imaging setup. When placed one focal length away from the imaging system it transforms momentum after Lens 3, which reflects the position of the effective imaging plane, onto the position distribution at the imaging setup. This is the configuration for $x$-Bohm measurement where post-selection is performed on position. Alternatively, Lens 3 could be placed at $f_2+f_3$ away from Lens 2. The resulting transformation with all three lenses and back propagation combined is 
\begin{equation}
    \left(
    \begin{array}{cc}
     0 & -f_1 \\
     \frac{1}{f_1} & -\frac{f_1+y}{f_1} \\
    \end{array}
    \right).
\end{equation}
Note the absence of $f_2$ in the equation. This is due to the fact that Lens 2 and Lens 3 are identical and they are configured such that a one-to-one telescope is formed. The transformation effectively places the focus of Lens 1 onto the imaging setup, performing a momentum post-selection for measurements in $p$-Bohm.

\section{Calibration}\label{sec:calibrationsec}
Our calibrartion proceedures can be split into two parts. The calibration of our lens system, where the position of the lenses is calibrated with the effective propagation distance, and the calibration of the weak measurement, where the strength of the weak measurement is determined.
\subsection{Lens System}
\begin{figure}
    \centering
    \includegraphics[width=0.5\columnwidth]{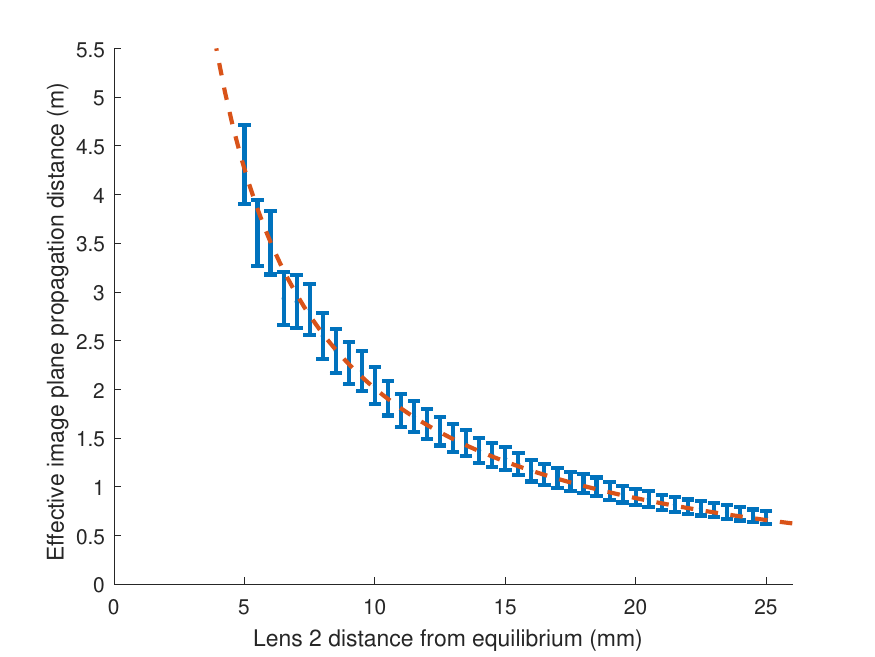}
    \caption{Calibration of the position of the second lens from the left and the corresponding propagation distance of the effective image plane. The origin position of the second lens is the place where far field is projected onto the camera. The blue vertical bars indicate the calibrated distance with uncertainty, where the uncertainty mostly originates from the beam profile of the individual slit not being perfectly Gaussian. The orange dotted line is the fitted curve of the calibration data.}
    \label{fig:DistanceCalibration}
\end{figure}

The effective propagation distance for the lens system is determined by comparing the ratio between the beam waist and slit center separation to to the same ratio if the beam is propagated in free space. Since the beam from the double slit is generated by a displaced triangular Sagnac, it is guaranteed that the slit separation in free space is constant, which in our case is $s= 2$\,mm. Moreover, the individual beams would also diverge, with waist increasing according to $w(z)=w_0\sqrt{1+(z/z_r)^2}$, where $z_r=1.04\,$m is the Rayleigh range. The ratio is thus, as a function of effective propagation distance, 
\begin{equation}\label{eq:lens_cali_ratio}
    R(T)=w_0\sqrt{1+(z/z_r)^2}/s.
\end{equation}
By measuring the same ratio under the lens system and applying inversion to equation (\ref{eq:lens_cali_ratio}), the effective propagation distance could be found.
The calibration of lens 2 is done by noting the magnification of the setup given the position of lenses. 
By measuring the waist of the beam from an individual slit, as well as the distance between the centroid of individual beams from the two slits, the effective propagation distance and magnification can be calculated. The results of the lens calibration is shown in Figure~\ref{fig:DistanceCalibration}.

\subsection{Weak measurement calibration}
\begin{figure}
    \centering
    \includegraphics[width=0.5\columnwidth]{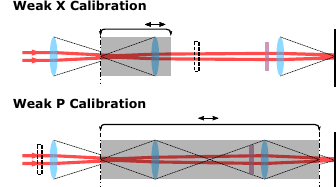}
    \includegraphics[width=0.5\columnwidth]{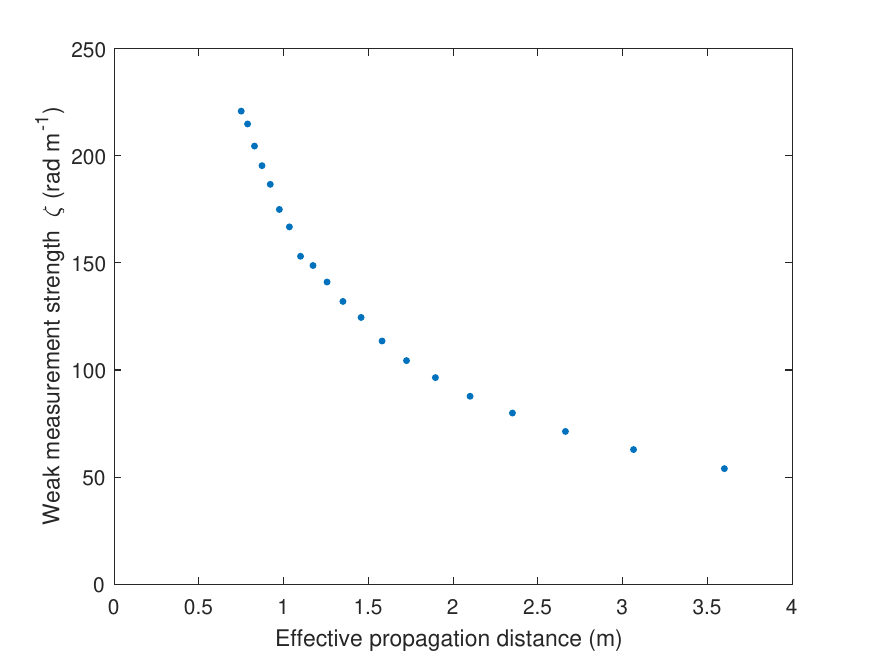}
    \caption{Illustration of calcite location for calibration of weak measurement strength $\zeta$. 
    Top (middle) figure corresponds to the setup where we calibrate the strength of weak position (momentum) measurement, where position (momentum) of the effective plane is measured both weakly and strongly.
    While the strength of weak momentum measurement is constant ($\zeta=134.49\pm 0.13$), the weak position measurement strength is a function of effective propagation distance and is plotted in the bottom figure. 
    Error bars, mostly due to the precision of the translation stage, are too small to show.
    }
    \label{fig:Strength_of_weakx}
\end{figure}

Calibration of the strength of the weak measurement $\zeta$ is determined by performing weak measurement and post-selection on the same variable, where $\omega$ and $\xi$ in Eq.~(\ref{eq:Holland}) are both set to be either position or momentum when calibrating for $x$-Bohm or $p$-Bohm, respectively. 
The corresponding setup and calibration results can be found in Figure \ref{fig:Strength_of_weakx}.

\end{document}